\documentclass[12pt,twoside]{article}
\usepackage{latexsym,amsmath,amssymb,amsthm,amsfonts,graphicx}
\usepackage{natbib}
\bibpunct{(}{)}{;}{a}{,}{,}

\def\,{\mskip 3mu} \def\>{\mskip 4mu plus 2mu minus 4mu} \def\;{\mskip 5mu plus 5mu} \def\!{\mskip-3mu}
\def\dispmuskip{\thinmuskip= 3mu plus 0mu minus 2mu \medmuskip=  4mu plus 2mu minus 2mu \thickmuskip=5mu plus 5mu minus 2mu}
\def\textmuskip{\thinmuskip= 0mu                    \medmuskip=  1mu plus 1mu minus 1mu \thickmuskip=2mu plus 3mu minus 1mu}
\textmuskip
\def\beq{\dispmuskip\begin{equation}}    \def\eeq{\end{equation}\textmuskip}
\def\beqn{\dispmuskip\begin{displaymath}}\def\eeqn{\end{displaymath}\textmuskip}
\def\bqa{\dispmuskip\begin{eqnarray}}    \def\eqa{\end{eqnarray}\textmuskip}
\def\bqan{\dispmuskip\begin{eqnarray*}}  \def\eqan{\end{eqnarray*}\textmuskip}

\newtheorem{theorem}{Theorem}

\newenvironment{keywords}{\centerline{\bf\small
Keywords}\begin{quote}\small}{\par\end{quote}\vskip 1ex}

\newtheorem{myexample}[theorem]{Example}

\def\paradot#1{\vspace{1.3ex plus 0.7ex minus 0.5ex}\noindent{\bf\boldmath{#1.}}}

\def\fr#1#2{{\textstyle{#1\over#2}}}

\def\SetR{I\!\!R}
\def\SetN{I\!\!N}

\def\B{\{0,1\}}
\def\E{{\bf E}}                         
\def\P{{\rm P}}                         

\def\v{\boldsymbol}

\def\b{\beta}

\def\B{{\cal B}}
\def\D{{\cal D}}
\def\F{{\cal F}}
\def\M{{\cal M}}

\def\H{{\cal H}}

\def\X{{\cal X}}
\def\Y{{\cal Y}}
\def\II{{\cal I}}

\def\crit{\text{\rm crit}}

\def\LR{\text{\rm LR}}

\def\pen{\text{\rm pen}}
\def\arg{\text{\rm arg}}
\def\Loss{\text{\rm Loss}}

\begin{document}

\title{\bf\Large\hrule height5pt \vskip 4mm
Model Selection by Loss Rank for Classification and Unsupervised Learning
\vskip 4mm \hrule height2pt}

\author{\\
{\bf Minh Ngoc Tran\footnote{This work was partially written while the author was visiting the RSISE@ANU in September, 2010.}}\\[3mm]
\normalsize Department of Statistics and Applied Probability \\
\normalsize National University of Singapore \ \ \
\normalsize \texttt{ngoctm@nus.edu.sg} \\[1ex]
\and \\
{\bf Marcus Hutter%
}\\[3mm]
\normalsize RSISE$\,$@$\,$ANU and SML$\,$@$\,$NICTA,
\normalsize Canberra, ACT, 0200, Australia \\
\normalsize \texttt{marcus@hutter1.net \ \ \ www.hutter1.net} \\
}%

\maketitle

\begin{abstract}\noindent
\cite{Hutter:07} recently introduced the loss rank principle (LoRP)
as a general-purpose principle for model selection.
The LoRP enjoys many attractive properties and deserves further investigations.
The LoRP has been well-studied for regression framework in \cite{Hutter:10}.
In this paper, we study the LoRP for classification framework, 
and develop it further for model selection problems in unsupervised learning 
where the main interest is to describe the associations between input measurements,
like cluster analysis or graphical modelling.
Theoretical properties and simulation studies are presented. 
\end{abstract}

\begin{keywords}
Classification,
graphical models,
loss rank principle,
model selection.
\end{keywords}

\newpage
\section{Introduction}\label{secIntro}
\paradot{Model selection}
Model selection is an important problem in machine learning and statistics.
Typically, model selection can be regarded as the question of choosing the right model complexity.
The maximum likelihood principle (MLP) breaks down
when one has to select among a set of nested models,
because then the MLP always selects the biggest model (w.r.t. inclusion).
Overfitting is a serious problem in structural learning from data.
Much effort has been put into developing model selection criteria that can avoid overfitting.
The most popular ones are probably AIC \citep{Akaike:73},
the BIC \citep{Schwarz:78}, the $C_p$ \citep{Mallows:73},
the MDL \citep{Rissanen:78}, cross-validation \citep{Allen:74,Craven:79}
and criteria based on Rademacher complexities \citep{Koltchinskii:01,Bartlett:02}. 
The reader is referred to \cite{Shao:96,Hutter:10} for comparisons of/some of these criteria. 
The loss rank principle (LoRP) introduced recently in \cite{Hutter:07,Hutter:10}
is another contribution to the model selection literature.
The LoRP, as it is named, is a general-purpose principle for model selection
rather than a specific criterion.
The LoRP can be regarded as a guiding principle for deriving model selection criteria 
that can avoid overfitting.
It has the advantage of always giving answers,
even in cases where the other criteria can not be used.

\paradot{The loss rank principle}
Consider the problem of selecting a model among a given set of models $\M$
achieving some kind of optimality properties.
The main goal of the LoRP is to establish a selection criterion 
that is able to specify a parsimonious model that fits the data not too bad.
General speaking, the LoRP consists in the so-called {\em loss rank} of a model
defined as the number of other (fictitious) data that fit the model better than the actual data,
and the model selected is the one with the smallest loss rank.

We now briefly present the LoRP developed in \cite{Hutter:07} and \cite{Hutter:10} for supervised learning settings.
In supervised learning, the data is categorized into {\em input} and {\em output},
and the main interest is to develop a model for predicting output based on input. 
Let $D=(\v x,\v y)=\{(x_1,y_1),...,(x_n,y_n)\}\in(\X\times \Y)^n$ be the (actual) training data set
with $\v x=(x_1,...,x_n)$ are inputs and $\v y=(y_1,...,y_n)$ are (disturbed) outputs.
Suppose that we use a model $M\in\M$ to fit the data $D$, 
e.g., $M$ is a linear regression model with $d$ covariates, 
or $M$ is a $k$-nearest neighbors regression model. 
Imagine that in experiment situations we can conduct the experiment many times 
with fixed design points $\v x$. 
We then would get many other (fictitious) outputs $\v y'$. 
Let $\Loss_M(\v y|\v x)$ be the empirical loss associated with a certain loss function 
when using a model $M\in\M$ to fit the data set $(\v x,\v y)$.
For instance, $\Loss_M(\v y|\v x)$ can be the least squares error, or
the negative maximum log-likelihood when a sampling distribution is assumed.
The loss rank of model $M$ then is defined as
\beq\label{lossrank}
\LR(M|D):=\mu\left\{\v y'\in\Y^n:\Loss_M(\v y'|\v x)\leq \Loss_M(\v y|\v x)\right\}
\eeq
with some measure $\mu$ on $\Y^n$.
For example, $\mu$ can be the counting measure if $\Y$ is discrete,
the usual Lebesgue measure on $\SetR^n$ if $\Y=\SetR$,
or some empirical probability measure (see Sections below).
Intuitively, the loss rank is large for too flexible models fitting $D=(\v x,\v y)$ well 
{\em and} also for too rigid models that fit $D$ not well
(in both cases the model fits many other $D'=(\v x,\v y')$ better).
For example, consider the polynomial regression problem where $(x_i,y_i)\in\SetR^2$,
a higher order polynomial would fit $D$ well
and also fit many other data $D'$ well,
thus resulting in a large loss rank. 
It was argued in \cite{Hutter:07} and \cite{Hutter:10} that 
minimizing the loss rank is a suitable model selection criterion
which trades off the quality of fit with the model flexibility.

The LoRP has been well studied for regression with continuous response \citep{Hutter:10}.
With continuous data and under squared loss, the loss rank has a closed form
and many optimality properties of the LoRP have been pointed out.
For example, the LoRP (i) is model selection consistent in some special cases;
(ii) reduces to Bayesian model selection in linear basis function regression with Gaussian prior;
(iii) has a minimum description length interpretation   
(interested readers are referred to \cite{Hutter:10} for the details).
Furthermore, the LoRP in supervised learning settings has been proven efficient 
in some specific applications.
\cite{Tran:09} demonstrated the use of LoRP for
selecting the ridge parameter in ridge regression,
while it was shown in \cite{Tran:10} that 
shrinkage parameters in regularization procedures like Lasso \citep{Tibshirani:96} or SCAD \citep{Fan:01}
selected by the LoRP enjoy good statistical properties.

The LoRP seems to be a promising principle with a lot of potential,
leading to a rich field.
We would like to emphasize that the LoRP should be regarded as a guiding principle
which in specific applications helps to derive model selection criteria that can avoid overfitting. 
This paper continues our investigation of the LoRP as a 
general-purpose procedure for model selection. 
We first study the LoRP for classification framework where the response is discrete.
Based on the LoRP, we derive a model selection criterion for classification 
and show that minimizing the criterion is asymptotically equivalent to     
minimizing an ideal criterion which is only known when the population distribution is known.

Second, we develop the LoRP for model selection in unsupervised learning settings.
This unsupervised learning LoRP then is studied by means of simulation
in two specific applications: selection of number of clusters in cluster analysis and 
model selection in graphical modelling.
The simulation shows that the model selection criteria derived from the LoRP work well and are competitive to existing ones. 

We end this introduction section by listing some attractive properties of the LoRP. 
The LoRP
\begin{itemize}\parskip=0ex\parsep=0ex\itemsep=0ex
\item always gives answers;
\item does not require insight into the inner structure of the problem; 
\item does not require any explicit setting of the stochastic noise structure, 
i.e. no assumption of sampling distribution is needed;
\item would work with any loss function.   
\end{itemize}

\section{Model Selection by Loss Rank for Classification}\label{secClass}
We consider in this section the model selection problem in a (binary) classification framework.
Let $D=\{(X_1,Y_1),...,(X_n,Y_n)\}$ be $n$ independent realizations of random variables $(X,Y)$,
where $X$ takes on values in some space $\X$ and
$Y$ is a $\{0,1\}$-valued random variable.
We assume that these pairs are defined on a probability space $(\Omega,\Sigma,\P)$ with $\Omega=(\X\times\Y)^n$.
We are interested in constructing a predictor $t:\X\to\{0,1\}$ that predicts $Y$ based on $X$.
The performance of the predictor $t$ is ideally measured by the prediction loss
\beq\label{pre.loss}
\P\gamma(t)=\P(I_{Y\not=t(X)})=\P(Y\not=t(X))
\eeq
where $\gamma(t)(x,y):=I_{y\not=t(x)}$ is called the contrast function.
Hereafter, for a measure $\mu$ and a $\mu$-integrable function $f$,
we denote the integral $\int fd\mu$ by $\mu f$ or $\mu(f)$.

Ideally, we want to seek an optimal predictor $s$
that minimizes $\P\gamma(t)$ over all measurable $t:\X\to\{0,1\}$.
However, finding such a predictor is impossible in practice
because the class of all measurable functions $t:\X\to\{0,1\}$ is huge and typically not specified.
Instead, we may restrict to some small class of predictors $\F$.
A question arises immediately here: how small should the class $\F$ be?
A too small $\F$ may lead to an unreasonable prediction loss,
while finding an optimizer in a too large $\F$ may be an impossible task.
Therefore the class/model $\F$ itself must be selected as well
(the terms {\it class} and {\it model} will be used interchangeably).
In this paper, we are interested in the model selection problem
in which we would like to find a good model (in a sense specified later on)
in a given set of models $\{\F_m,\ m\in\M\}$.

The unknown prediction loss \eqref{pre.loss} is often estimated by the empirical risk
\beq\label{emp.risk}
\P_n\gamma(t)=\fr1n\sum_1^nI_{Y_i\not=t(X_i)}
\eeq
where $\P_n$ is the empirical measure based on data $D$
\beqn
\P_n=\fr1n\sum_1^n\delta_{(X_i,Y_i)}
\eeqn
with $\delta_x$ denotes the Dirac measure at $x$.
For a class $\F_m$, one may seek a function $\hat t_m$
minimizing $\P_n\gamma(t)$ over $t\in\F_m$.
Unfortunately, it is well-known that such a method leads to overfitting:
the larger $\F_m$, the smaller the empirical risk $\P_n\gamma(\hat t_m)$.
Consequently, the selected model is always the biggest one.
This leads to the idea of accounting for the model complexity,
in which we select a model $\hat m$ that minimizes the sum of the empirical risk
and a penalty term taking the model complexity into account.

Because $\P_n\gamma(t)$ underestimates $\P\gamma(t)$, a well-known regularized criterion
for model selection is to penalize the approximation on $\F_m$ of the prediction loss by the empirical risk
(see, e.g., \cite{Koltchinskii:01,Fromont:07,Arlot:08})
\beq\label{idealcriterion}
\crit_n(m)=\P_n\gamma(\hat t_m)+\sup_{t\in \F_m}(\P-\P_n)\gamma(t).
\eeq
The second term, denoted by $\pen_n(m)$, is a natural measure of the complexity of class $\F_m$,
which measures the accuracy of empirical approximation on class $\F_m$.
Then, the model to be selected is $m_n=\arg\min_m\{\crit_n(m)\}$.
For simplicity, we assume throughout the paper that $m_n$ is uniquely determined. 

In practice, $\P$ is unknown and so is $\pen_n(m)$.
One has to estimate $\pen_n(m)$.
Many methods have been proposed to estimate this theoretical penalty: VC-dimension \citep{Vapnik:71},
Rademacher complexities \citep{Koltchinskii:01,Bartlett:02},
resampling penalties \citep{Fromont:07,Arlot:08}.
All of these methods give upper bounds for $\pen_n(m)$.
The performances of the methods are measured in terms of oracle inequalities.
The sharper the estimate is, the better the performance is.
These methods often works well in practice but are not without problems.
For example, the VC-dimension is often unknown and needs to be estimated by another upper bound,
Rademacher complexities are often criticized to be too large
(the local Rademacher complexities \citep{Bartlett:05, Koltchinskii:06} have been introduced to
overcome this drawback, however the latter still suffer from the hard-calibration problem
because they involve unknown constants).

In this section, based on the LoRP, we propose a criterion to estimate the model $m_n$ directly, {\it not $\pen_n$}.
Instead of giving an upper bound for $\pen_n(m)$, we directly estimate $m_n$
by minimizing a criterion over models $m\in \M$.
Minimizing the criterion is asymptotically equivalent to minimizing $\crit_n(m)$
with probability 1 (Theorem \ref{theo}).

In Section \ref{secMainResult},
the suggested criterion is derived and its model consistency is proven.
In Section \ref{secCalibration}, we discuss the implementation
and carry out a numerical example to demonstrate the criterion and compare it to other methods.

\subsection{The loss rank criterion}\label{secMainResult}
The LoRP, as it is named, is a guiding principle rather than a specific selection criterion.
When it comes to apply in a specific context,
a suitable choice of measure $\mu$ in \eqref{lossrank} is needed.
For continuous data cases, using the usual Lebesgue measure in $\SetR^n$ 
leads to a closed form of loss rank and meaningful results \citep{Hutter:10}. 
In our current context of the binary classification,
some suitable probability measure on $\Y^n=\{0,1\}^n$
should be used to define the loss rank.
To formalize this, we define the loss rank of a model
as the probability that a randomly resampled sample fit the model better than the actual sample.
This definition of the loss rank makes it not only possible to estimate the loss rank
but also makes use of the available theory of resampling to justify the method.

We now formally define the loss rank.
Let $r_i,\ i=1,...,n$ be $n$ independent Rademacher random variables,
i.e. $r_i$ takes on values either $-1$ or $1$ with probability $1/2$.
The $r_i$'s are assumed to be independent of $D$.
Let $Y_i':=\fr{1+r_i}{2}-r_iY_i$, i.e. we flip the value/label of $Y_i$ with probability $1/2$.
The loss rank of model $m$ is defined as
\beq\label{LoRPdefinition}
\LR_n(m)\equiv\LR_n(\F_m):=\P_R(\inf_{t\in\F_m}\fr1n\sum_1^nI_{Y_i'\not=t(X_i)}\leq \P_n\gamma(\hat t_m)|D)
\eeq
where $\P_R(.|D)$ means the conditional probability w.r.t.\ the Rademacher sequence given data $D$.
Intuitively, the empirical risk based on the actual $D$ would be small for a too flexible class $\F_m$,
but many resamples $D'$ would then also result in small empirical risk,
which leads to a large loss rank $\LR_n(m)$.
Therefore, minimizing the loss rank helps avoid overfitting.
Also, a too rigid $\F_m$ fitting $D$ not well would lead to a large loss rank as well.
Thus, the loss rank defined in \eqref{LoRPdefinition}
is a suitable criterion for model selection which trades off between the fit (empirical risk)
and the model complexity.

$\LR_n(m)$ is directly estimable by a simple Monte Carlo algorithm (see the next section).
Then the selected model will be $\hat m_\LR=\arg\min_{m\in\M}\LR_n(m)$.
We name this method the loss rank (LR) criterion.

\paradot{Optimality property}
We now discuss the model consistency of the LR criterion
by using the modern theory of empirical processes
(see, e.g., \cite{vdV:96}).
To avoid dealing with difficulties of non-measurability in empirical process theory,
we as usual assume that for each $m\in\M$, class $\F_m$ is countable.
We need the following regularity condition:
\begin{itemize}
\item[(C)] $\D_m=\{\gamma(t),t\in\F_m\}$, $m\in\M$ are Donsker classes.
\end{itemize}
Recall that a function class $\D$ is called a Donsker class if $\sqrt{n}(\P_n-\P)f$ converges in probability
to $N(0,\P(f-\P f)^2)$ uniformly in $f\in\D$.
This, together with another condition that $\P\left(\sup_{f\in\F}|f-\P f|^2\right)<\infty$
(which is automatically satisfied in our context because $\gamma(t)\leq1$ for every predictor $t$)
are essential in order for the weak convergence of empirical processes to hold \cite[Chapter 3]{vdV:96}.
These are also two essential conditions in order for Efron's bootstrap to be asymptotically valid \citep{Gine:90} (see also \cite{vdV:96}).
\begin{theorem}\label{theo}
Under Assumption (C),
minimizing $\LR_n(m)$ over $m\in\M$ is asymptotically equivalent to
minimizing the ideal criterion $\crit_n(m)$ with probability 1,
i.e. $\hat m_\LR$ is a strong consistent estimate
of $m_n$.
\end{theorem}
On one hand, LR criterion is closely related to penalized model selection based on Rademacher complexities.
As being realized by \cite{Lozano:00}, a very large model which generally contains a predictor
predicting correctly most of randomly generated labels results in a large Rademacher penalty.
While a very large model will result in a large loss rank which is defined as
the probability that a randomly relabeled sample behaves better than the actual sample.
On the other hand, LR criterion is quite different from Rademacher complexities model selection.
While Rademacher complexities give upper bounds for the ideal penalty $\pen_n(m)$,
LR criterion offers a way to directly estimate the ideal model $m_n$.

\begin{proof}[Proof of the theorem]
By $Y_i':=\fr{1+r_i}{2}-r_iY_i$, it's easy to see that $I_{Y_i'\not=t(X_i)}=I_{r_i=1}-r_iI_{Y_i\not=t(X_i)}$, therefore
\beq\label{pt1}
\inf_{t}\fr1n\sum_1^nI_{Y_i'\not=t(X_i)}=\fr1n\sum_1^nI_{r_i=1}-\sup_t\fr1n\sum_1^nr_iI_{Y_i\not=t(X_i)}.
\eeq
Moreover,
\beq\label{pt2}
\fr1n\sum_1^nr_iI_{Y_i\not=t(X_i)}=\fr1n\sum_1^nI_{Y_i\not=t(X_i)}-\fr1n\sum_1^n(1-r_i)I_{Y_i\not=t(X_i)}=\P_n\gamma(t)-\P_n^R\gamma(t)
\eeq
where $\P_n^R:=\fr1n\sum W_i\delta_{(X_i,Y_i)}$ with $W_i:=1-r_i\sim 2\text{Binomial}(1,1/2)$
is the {\em weighted bootstrap empirical measure}. From \eqref{pt1}-\eqref{pt2} and \eqref{LoRPdefinition}, we have
\beqn
\LR_n(m)=\P_R\Big(\sup_{t\in\F_m}(\P_n-\P_n^R)\gamma(t)\geq\fr1n\sum_1^nI_{r_i=1}-\P_n\gamma(\hat t_m)\big|D\Big).
\eeqn
The key point in the proof is the result of weak convergence
of the weighted bootstrap empirical processes.
The result states that, under Assumption (C),
the difference between the conditional law of $\P_n-\P_n^R$ given data $D$ and the law of $\P-\P_n$
converges to zero almost surely (see \cite[p.346]{vdV:96}).
More formally, let $\hat G_n=\P_n-\P_n^R$ and $G_n=\P-\P_n$,
and let $l^\infty(\D_m)$ be the space of all bounded functions from $\D_m$ to the real set $\SetR$
($\hat G_n$ and $G_n$ are random elements in $l^\infty(\D_m)$).
Then
\beqn
|\E_Rh(\hat G_n)-\E h(G_n)|\to0,\;\;\P-\text{almost surely}
\eeqn
for every continuous, bounded function $h:l^\infty(\D_m)\to\SetR$. 

Therefore, by the continuous mapping theorem with notice that $\fr1n\sum_1^nI_{r_i=1}\to1/2$ a.s., we have $\P$-almost surely
\bqan
&\Big|\P_R\Big(\sup_{t\in\F_m}(\P_n-\P_n^R)\gamma(t)\geq\fr1n\sum_1^nI_{r_i=1}-\P_n\gamma(\hat t_m)\big|D\Big)\\
&-\P\Big(\sup_{t\in\F_m}(\P-\P_n)\gamma(t)\geq\fr12-\P_n\gamma(\hat t_m)\Big)\Big|\to0.
\eqan
Thus, as n is sufficiently large
\beqn
\LR_n(m)=\P\left(\sup_{t\in\F_m}(\P-\P_n)\gamma(t)\geq\fr12-\P_n\gamma(\hat t_m)\right)=\P(\crit_n(m)\geq\fr12)\;\; \text{w.p.1.}
\eeqn

For simplicity, suppose now that $\LR_n(m)$ has a unique minimum at $\hat m_\LR$.
If $\hat m_\LR\not=m_n$, $\P(\crit_n(m_n)\geq\fr12)>\P(\crit_n(\hat m_\LR)\geq\fr12)$.
On the other hand, $\crit_n(m_n)<\crit_n(\hat m_\LR)$ by the definition of $m_n$, 
so $\P(\crit_n(m_n)\geq\fr12)\leq\P(\crit_n(\hat m_\LR)\geq\fr12)$.
The contradiction implies $\hat m_\LR=m_n$ w.p.1.
\end{proof}

\subsection{Implementation and Simulation}\label{secCalibration}

\paradot{Implementation}
The loss rank $\LR_n(m)$ can be easily estimated
by a simple Monte Carlo algorithm as follows:
\begin{itemize}
\item[1.] $\hat\LR_n(m)\leftarrow 0$.
\item[2.] Toss a fair coin $n$ times and define
\beqn
Y_i'=\begin{cases}
Y_i,&\text{head occurs at $i$-th time}\\
1-Y_i,&\text{tail occurs at $i$-th time}
\end{cases},\ i=1,2,...,n.
\eeqn
If $\inf_{t\in\F_m}\fr1n\sum_1^nI_{Y_i'\not=t(X_i)}\leq \P_n\gamma(\hat t_m)$ then $\hat\LR_n(m)\leftarrow \hat\LR_n(m)+1/B$.
\item[3.] Repeat step 2, $B$ times.
\end{itemize}
The theoretical justification for this algorithm is the law of large numbers: $\hat\LR_n(m)\to\LR_n(m)\ a.s.$ as $B\to\infty$.
In the following simulation, $B$ is taken to be 200.
From our experience, the results do not change much if a larger $B$ is used.

\paradot{A numerical example}
We now demonstrate the method by a simple example of a piecewise constant classifier with $2^m$ segments.
and compare it to model selection based on Rademacher complexities.
Consider the intervals model selection problem which was described by \cite{Fromont:07} (see also, \cite{Lozano:00,Bartlett:02}).
Given a number $N\in\SetN$, let $\X=\{1,2,...,2^N\}$.
For $u,v\in\SetN, u\leq v$, denote by $\SetN[u,v]$ the set of integers in interval $[u,v]$.
For an integer number $m,\ 1\leq m\leq N$, let
\beqn
\F_m=\left\{t:\X\to\{0,1\},t=\sum_{k=1}^{2^m}c_kI_{\SetN[(k-1)2^{N-m}+1,k2^{N-m}]},c_k\in\{0,1\},k=1,...2^m\right\}
\eeqn
be the set of piecewise constant functions defined on $\X$ and
taking on values $\{0,1\}$ with possible jumps at $k2^{N-m},\ k=1,...,2^m-1$.

For a given $m_0,\ 1\leq m_0\leq N$, let $S_0$ be the set of odd-numbered segments:
\beqn
S_0=\bigcup_{k=2p+1,\ p=0,1,...,2^{m_0-1}-1}\SetN[(k-1)2^{N-m_0}+1,k2^{N-m_0}].
\eeqn
Let $X$ be a uniformly distributed random variable on $\X$
and $Y$ be a $\{0,1\}$-valued random variable defined as
\beqn
\P(Y=1|X\in S_0)=\fr12+h,\;\;\text{and}\;\;\P(Y=1|X\notin S_0)=\fr12-h
\eeqn
where $h\in(1,\fr12)$ is called the margin parameter.
We now have a model selection problem with $N$ candidate models $\{\F_m,\ m\in\M=\{1,...,N\}\}$
and the optimal predictor $s(x)=I_{S_0}(x)\in \F_{m_0}$ belongs to one of them.
We are interested in identifying the true model $m_0$.
The advantage of the intervals model selection problem is that
it is very easy to compute for each $m\in\M$
\beqn
\P_n\gamma(\hat t_m)=\inf_{t\in \F_m}\fr1n\sum_{i=1}^nI_{Y_i\not=t(X_i)}\;\text{and}\;\sup_{t\in \F_m}\fr1n\sum_{i=1}^nr_iI_{Y_i\not=t(X_i)}.
\eeqn
The reader is referred to \cite{Fromont:07} for the details.

We compare LR criterion to another criterion based on Rademacher complexities
which is taken following \cite{Fromont:07} to be
\beqn
\crit_\text{RC}(m)=\P_n\gamma(\hat t_m)+\pen_\text{RC}(m)\;\text{with}\;\pen_\text{RC}(m)=E(\sup_{t\in\F_m}\fr1n\sum_{i=1}^nr_iI_{Y_i\not=t(X_i)}|D)
\eeqn
We shall call this the Rademacher complexity (RC) criterion.
In our experiment, Rademacher complexities $\pen_\text{RC}(m)$ are estimated also by 200 Monte Carlo simulations.

\begin{figure*}
\centerline{\includegraphics[width=1.3\textwidth,height=.6\textwidth]{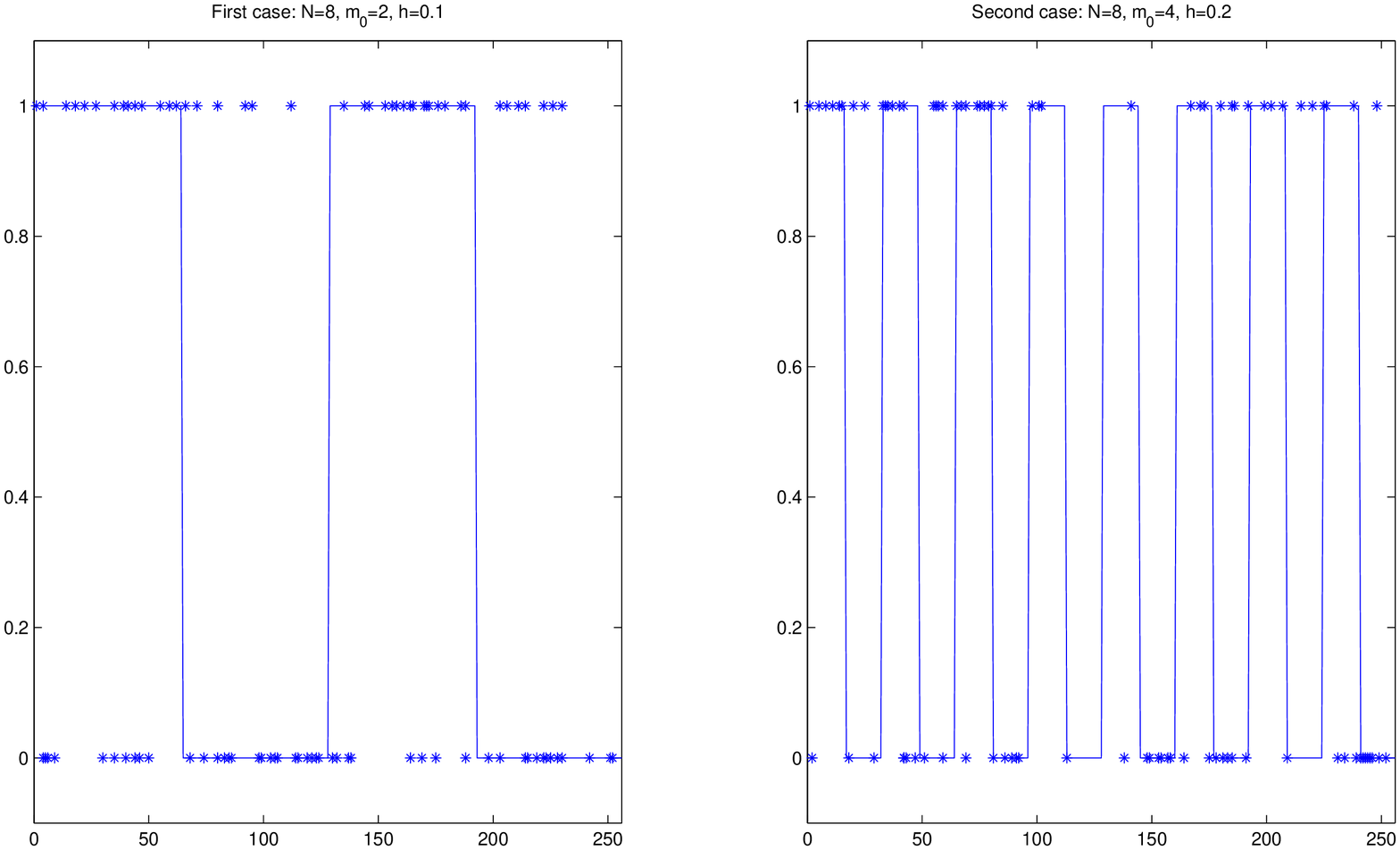}}
\caption{\label{DataPlot}
The plots of true functions and data for two cases.}
\end{figure*}
\begin{figure*}
\centerline{\includegraphics[width=1.3\textwidth,height=.6\textwidth]{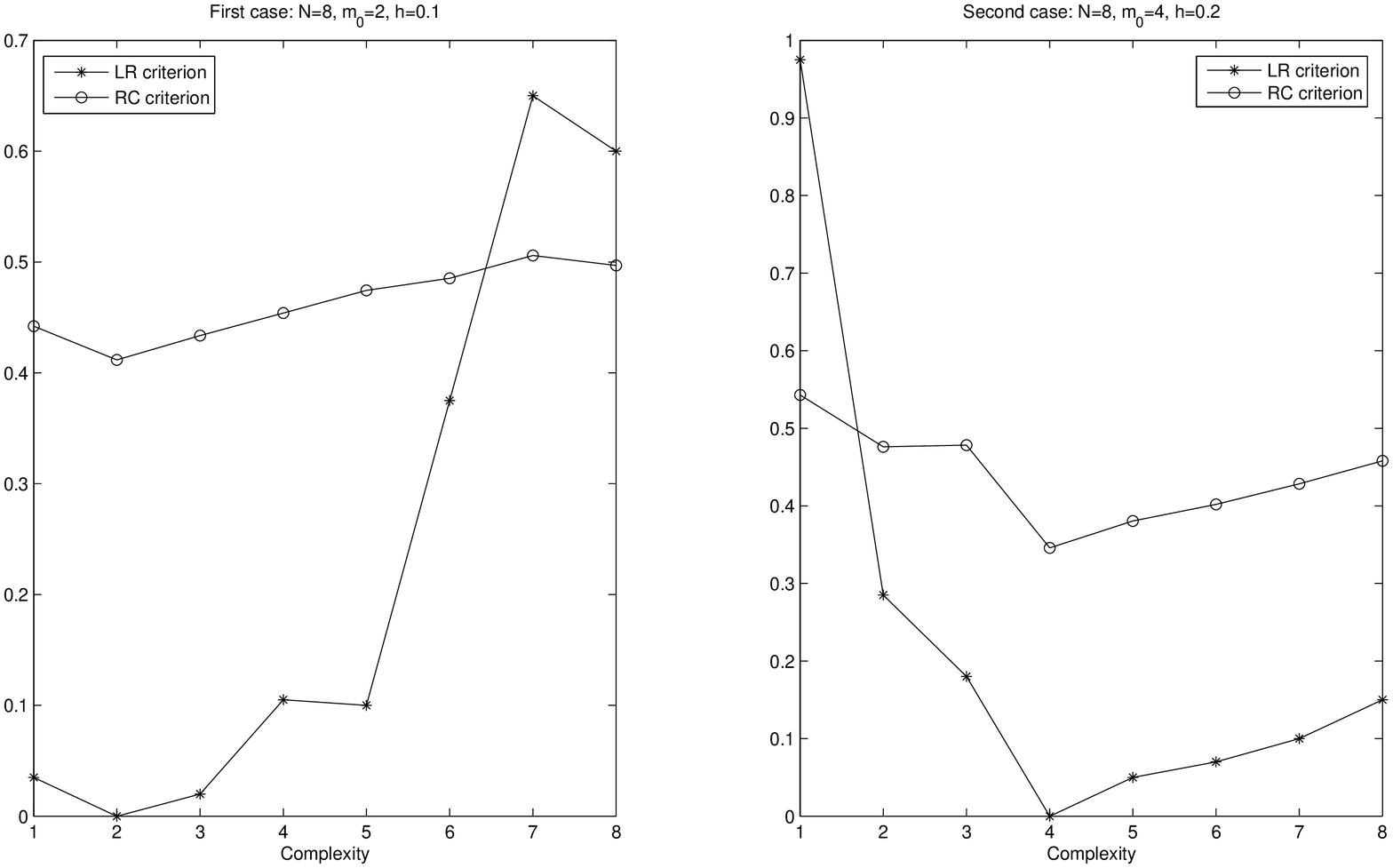}}
\caption{\label{BLoRP}
The plots of LR criterion and Rademacher complexity criterion.}
\end{figure*}

Figure \ref{DataPlot} plots true functions and observation data (with $n=100$) for two cases:
first with $N=8,\ m_0=2,\ h=.1$, then $N=8,\ m_0=4,\ h=.2$.
These pictures show how hard it is to decide intuitively
what the true model is. 
Figure \ref{BLoRP} plots LR criterion and RC criterion.
Both criteria identify the true model in both cases.

Table 1 presents the proportions of correct identification
over $100$ replications for each of $16$ cases with various sample sizes
$n=50,\ 100,\ 200,\ 300$ and noise levels $h=.05,\ .1,\ .2,\ .3$  ($m_0=4$).
It is shown that both criteria are model selection consistent 
as the proportions increases to 1 as $n$ and $h$ increase.
The simulation suggests that the LR criterion has an improvement over the RC criterion for large sample sizes.

\begin{table}[ht]
\centering 
\begin{tabular}{c|c|c|c||c|c|c|c}
$n$&$h$&LR criterion&RC criterion&$n$&$h$&LR criterion&RC criterion\\
\hline\hline
50&.05&.12&.13&200&.05&.23&.21\\
  &.1&.35&.35&  &.1&.67&.66\\
  &.2&.62&.64&  &.2&.99&.97\\
  &.3&.95&.97&  &.3&1&1\\
\hline
100&.05&.15&.15&300&.05&.30&.28\\
   &.1&.41&.41&    &.1&.78&.76\\
   &.2&.89&.90&    &.2&1&.99\\
   &.3&.98&.98&    &.3&1&1
\end{tabular}
\label{table1} 
\caption{Proportions of correct identification of LR and RC criterion for various $n$ and $h$.} 
\end{table}

\section{The LoRP for unsupervised learning}\label{secUnsup}
The LoRP developed so far is for supervised learning settings only.
In supervised learnings, there are measurements called {\em inputs}
which are used to predict {\em outputs}.
Note that, in such settings, we have fixed the inputs $\v x$ in the definition of the loss rank,
and ``resample" only the outputs $\v y$.
This seems to have some physical interpretation in supervised learnings and more importantly 
leads to a closed form of loss rank in many cases \citep{Hutter:10}.
Such a way is not applicable to unsupervised learning settings where 
there is no outputs.
For example, in graphical modelling or cluster analysis,
the question of interest is to explore the associations between a set of input measurements.  
Fortunately, the basic reasoning of LoRP can be straightly extended to unsupervised learning.
It is worth recalling the key observation of the LoRP: too flexible models will fit the actual data well
and also fit fictitious/resampling data well (``fitting well" here means ``having a small empirical loss").
Let $\v x=(x_1,...,x_n)$ be the actual data set and $\Loss_M(x)$ be the empirical loss
when fitting data $\v x$ by a model $M$.
Assume that the empirical loss has the property that the more flexible $M$, the smaller $\Loss_M(\v x)$.
Let $\v x'$ be a resample from $\v x$ using some resampling scheme (e.g., boostrapping).
We can now define the loss rank of model $M$ as
\beqn
\#\{\v x':\Loss_M(\v x')\leq \Loss_M(\v x)\}.
\eeqn
This definition is easily understood intuitively but not very practical 
because the total number of resamples $\v x'$ is often huge or infinite.
To make it more practical, we can proceed as follows.
Let $\B$ be the set of $B$ resamples $\v x'$ from $\v x$.
The loss rank now can be defined as
\beq\label{lossun}
\hat\LR_B(M|\v x) = \frac{\#\{\v x'\in\B:\Loss_M(\v x')\leq \Loss_M(\v x)\}}{B}.
\eeq
Mathematically, let $\hat\P_n$ be the empirical probability measure of the resampling scheme
\citep{Efron:93,vdV:96}, we formally define the loss rank as
\beq\label{lossunf}
\LR(M|\v x) = \hat\P_n\{\v x':\Loss_M(\v x')\leq \Loss_M(\v x)\}.
\eeq
Clearly, the loss rank defined in \eqref{lossun} is an estimate of the one defined in \eqref{lossunf}.
In the next section, we will study the unsupervised LoRP by means of simulation.
The resampling scheme used is the popular bootstrap \citep{Efron:93}. 

\section{Simulation studies for unsupervised LoRP}\label{secGraph}
In this section, the unsupervised LoRP will be applied to 
selecting good models in graphical modelling
and selecting number of clusters in cluster analysis.
\subsection{LoRP for choosing number of clusters}
Cluster analysis \citep[Ch.14]{Hastie:05} is an important problem in unsupervised learning. 
The goal is to group a collection of objects into clusters
such that objects within each cluster are more closely related to each other 
than objects assigned to different clusters.
In some applications, the number of clusters $K$ may be known in advance
but in most cases $K$ is unknown and must be selected based on the data.
Popular methods for model selection such as AIC, BIC or coss-validation are not applicable here
(see, e.g., \cite{Hastie:05}, Ch.14).
Let $\v x=(x_1,...,x_n)$ be $n$ objects and $d(x_i,x_j)$ be the distance (or dissimilarity measure)
between $x_i$ and $x_j$.
Suppose that the $n$ objects $\v x$ have been clustered into $K$ clusters $C_1,...,C_K$
using some clustering algorithm (e.g., the K-means algorithm).
The natural loss is the {\em within-cluster} sum of dissimilarities
\beqn
W_K(\v x) = \frac12\sum_{k=1}^K\sum_{i,j\in C_k}d(x_i,x_j).
\eeqn 
When number of clusters $K$ increases, $W_K$ generally decreases.
Let $\B$ be the set of $B$ bootstrap resamples $\v x'$ from the actual data $\v x$,
we define the loss rank of using $K$ clusters as in \eqref{lossun} by
\beqn
\hat\LR_B(K|\v x)=\frac{\#\{\v x'\in\B:W_K(\v x')\leq W_K(\v x)\}}{B}.
\eeqn
The optimal $K$ selected by the LoRP will be $\hat K_\LR=\arg\min_K{\hat\LR}_B(K|\v x)$.

A popular method in the literature for selecting $K$ is the criterion proposed in \cite{Calinski:74}
\beqn
\text{CH}(K)=\frac{B_K/(K-1)}{W_K/(n-K)},
\eeqn
and the $K$ selected is the one maximizing this criterion.
Note that the CH criterion is not defined for $K=1$.
In the following simulation we compare the performance of the LoRP with that of the CH. 

\paradot{Simulation} 
We generate 2-dimensional datasets with various settings:
\begin{itemize}
\item 2 clusters, each with 50 observations, are generated
from 2-dimensional normal distributions $N(\mu,\sigma I)$ with $\mu=(0,0),\ (0,5)$ and $\sigma=1,2,3$.
\item 3 clusters, each with 50 observations, are generated
from 2-dimensional normal distributions $N(\mu,\sigma I)$ with $\mu=(0,0),\ (0,5),\ (5,0)$ and $\sigma=1,2,3$.
\item 4 clusters, each with 50 observations, are generated
from 2-dimensional normal distributions $N(\mu,\sigma I)$ with $\mu=(0,0),\ (0,5),\ (5,0),\ (5,5)$ and $\sigma=1,2,3$.
\end{itemize}
We measure the performance in terms of percentage that the true number of clusters is correctly identified,
over 100 replications for each setting.
The simulation result is summarized in Table \ref{tablecluster}.
It seems hard to compare the performance of the two methods.
While the CH outperforms the LoRP for ``easy" cases (small $\sigma$),
the LoRP outperforms the CH for ``hard" cases (large $\sigma$).
However, our main interest is not the quality of the LoRP in this particular example,
but to show that the unsupervised LoRP developed above as a general-purpose principle
works for selecting number of clusters.    
\begin{table}[ht]
\centering
\begin{tabular}{c|c|c|c}
$\#$ clusters&$\sigma$&CH&LR\\
\hline
2	&1	&1	&0.82\\
	&2	&1	&0.74\\
	&3	&1	&0.86\\
\hline
3	&1	&0.99	&0.84\\
	&2	&0.7	&0.45	\\
	&3	&0	&0.39\\
\hline
4	&1	&0.92	&0.56\\
	&2	&0.04	&0.38\\
	&3	&0	&0.50
\end{tabular}
\caption{Percentages of correct identification over 100 replications.}\label{tablecluster}
\end{table}

\subsection{LoRP for graphical modelling}
We study in this section the unsupervised LoRP for structural learning
in graphical modelling, mainly focus on Markov networks with discrete-valued vertices
(also called graphical log-linear modelling) \citep{Whittaker:90,Edwards:00}, 
but exactly the same idea would work for Bayesian networks as well. 

\paradot{Graphical modelling}
The basic idea of graphical modelling is to use graphs to represent
the independence structure among a set of variables.
A graph is a pair $G=(V,E)$ where the vertex set $V$ consists of a finite set of random variables
and the edge set $E$ represents the (conditional) independence relations between the r.v.'s in $V$.
For every $u,v\in V$, if $u$ and $v$ are conditionally independent given all the other variables in $V$
then the edge $(u,v)$ is {\it not} included in $E$.
In other words, a non-adjacent pair of vertices can be immediately interpreted 
as being conditionally independent given the rest.
Graphical modelling provides an efficient way to represent and communicate the conditional independence relations between a set of r.v.'s.   
We restrict ourselves to undirected graphs (also called Markov networks) in this paper, 
but the same idea can be directly adapted for directed ones (or Bayesian networks).

Most of the literature on graphical modelling is concerned with selecting an appropriate model to explain the data.
The most popular method is stepwise selection \citep{Whittaker:90,Edwards:00}
which starts at an initial base model and moves to next step  by including or excluding a single edge
until some termination criterion is fulfilled.
Stepwise selection is search-efficient but its main drawback is that it may get stuck in a local optimum \citep{Whittaker:90,Edwards:00}.
Furthermore, it is not easy to understand the statistical properties of the selected model.

Another criterion can be used for graphical model selection is the Bayesian information criterion (BIC) \citep{Schwarz:78}.
BIC of model $G$ has the form
\beqn
\text{BIC}(G)=-\log(\text{maximum likelihood under G})+\frac12(\#\text{free parameters of $G$})\log n 
\eeqn
It is well-known that BIC is asymptotically able to identify the true model (if it exists).
One may use AIC \citep{Akaike:73} as a selection criterion as well,
but AIC tends to select overfitted models.
AIC is optimal in terms of mean squared error loss \citep{Shibita:84},
however, this quantity is not well-defined in the graphical modelling context.
  
\paradot{Loss rank criterion}
Let $V$ be a set of $k$ discrete variables,
and for each $v\in V$ let $\II_v$ be the set of its possible values/levels.
The dataset of size $n$ is cross-classified by the levels of variables in $V$.
Let $\II=\otimes_{v\in V}\II_v$. The dataset is often conveniently given in the form of a contingency table 
with cell counts $\v n=\{n_i\}_{i\in\II}$ 
where $n_i$ is the observed number of observations cross-classified into cell $i$, $\sum n_i=n$.
The sampling distribution of $\v n$ is often assumed to be multinomial
\beqn
p(\v n|\v m)=\frac{n!}{\prod_{i\in\II}n_i!}\prod_{i\in\II}(\frac{m_i}{n})^{n_i},
\eeqn  
where $\v m=\{m_i\}_{i\in \II}$, $m_i$ are the expected numbers of observations falling into cells $i$ out of total $n$ observations, $\sum_i m_i=n$. 

Let $\{\hat m_i(G,\v n)\}$ be the MLE of $m_i$ under model $G$.
We define the empirical loss function resulting from fitting data $\v n$ by model $G$ as the negative maximum log-likelihood
(neglecting the constant terms depending only on $n$)
\beqn
\Loss_G(\v n): = -\sum_i\left[n_i\log(\hat m_i(G,\v n)) - \log(n_i!)\right]. 
\eeqn
This empirical loss is not a suitable measure for model selection,
because the larger the model $G$ (w.r.t. inclusion), the smaller the loss.
From \eqref{lossunf}, the loss rank of model $G$ is
\beq\label{GLoRP}
\LR_n(G): = \hat\P_n\left(\Loss_G(\v n')\leq\Loss_G(\v n)\right) 
\eeq
where $\hat\P_n$ denotes the bootstrap empirical measure \citep{Efron:93} 
and $\v n'=\{n_i'\}_{i\in\II}$ is a bootstrap resample from the actual data $\v n$.
The graph to be selected will be $\hat G_\LR=\arg\min\LR_n(G)$.
We call this strategy the loss rank criterion for graphical model selection.

Similar to the classification case, it is straightforward to estimate the loss rank \eqref{GLoRP}
by a simple Monte Carlo algorithm.
In the following simulation, we estimate $\LR_n(G)$ by an average over $B=200$ bootstrap resamples $\v n'$ from $\v n$.
From our own experience, the result does not change much if a larger number of replications is used.

Note that definition \eqref{GLoRP} is somewhat similar to definition \eqref{LoRPdefinition}
of the loss rank for classification. 
However, the proof technique in Theorem \ref{theo} seems not to apply here
because the derivations \eqref{pt1}-\eqref{pt2} in the proof are not valid anymore.
Instead, in the following we will evaluate the suggested strategy by means of simulation. 
A theoretical justification is left for the future work.

In order to help the reader grasp better how the LR criterion works,
we first present here a simple example where an exhaustive search over model space is possible.
For the case of large number of vertices $k$,
we will derive a genetic algorithm to
overcome the difficulty in searching over huge model spaces.

\paradot{A simple example}
We consider a simple example where the number of vertices is $k=3$,
and each variable takes on 3 values/levels.
The number of graphs then is $2^{\binom{k}{2}}=8$.
For a given sample size $n$, 100 datasets are generated from the ``true" model with formula $12/23$,
i.e., the first and third variable are conditionally independent given the second.
We evaluate the performance in terms of proportion of correct identification
over 100 replications.
Table \ref{table2} shows the performance of LR in comparison to that of BIC.
\begin{table}[ht]
\centering 
\begin{tabular}{c|ccccc}
$n$&200&500&1000&2000&5000\\
\hline
LR&.2&.7&.9&1&1\\
BIC&.05&.4&.7&.8&1
\end{tabular}
\caption{Proportions of correct identification of LR and BIC for various $n$} 
\label{table2}
\end{table}
The simulation result suggests that LR is superior to BIC.
This result is similar to the simulation result in \cite[Table 1]{Hutter:10} in which
it was also shown that the LoRP works better than BIC for model selection in linear regression.

\paradot{Graphical model selection with LR criterion and a genetic algorithm}
The main difficulty in graphical model selection is that
the number of models is increasing more than exponentially
as the number of vertices increases.
Model selection can be seen as a problem of searching for the optimal solution,
w.r.t. a certain selection criterion, over the model space.
A natural choice is to adapt genetic algorithms (GA) \citep{Holland:75,Mitchell:96} for searching over the model space.
This idea has been already taken in \citep{Poli:98}
who used AIC \citep{Akaike:73} as the selection criterion 
and proposed a genetic algorithm for model search.
Here, we adapt their genetic algorithm for model search and use the LR criterion as the selection criterion.

Genetic algorithms \citep{Holland:75,Mitchell:96} are widely used to search for optimization solutions
when the solution space is huge. 
The basic idea of GA is to mimic the evolutionary processes of creatures
in which they attempt to find better solutions to the given problem
by generating successive generations of individuals that are expected 
to be better suited to the environment than their ancestors.

Solutions are typically encoded by binary strings, called {\it chromosomes}.
Chromosomes are associated with a {\it fitness function}
and the problem is to find the fittest individual. 
The search space consists of all possible chromosomes,
which is typically infeasible to access every individuals.
A GA starts by generating an {\it initial population}
and proceed by applying in turn three operators: {\it selection, crossover} and {\it mutation}.
Selection operator randomly selects parents from the current population with 
probability being an increasing function of fitness to form a new population.
At this stage, another operator called {\it elitism} may be used,
in which a certain number of fittest individuals in the current population are 
directly inserted into the new population.
Offsprings are obtained by applying the crossover 
to pairs of parents with a probability of $p_c$ - a pre-fixed number in $[0,1]$.
The crossover typically consists in exchanging certain bits of two selected chromosomes.
Finally, the new generation is obtained by applying with a probability of $p_m$ the mutation operator 
which changes one or more bits of a chromosome.
The procedure is repeated until a termination criterion is satisfied.
A widely-used termination criterion is that the fittest does not change for the last, say $T$, iterations.
Some theoretical conditions to assure the convergence to the global optimal
were introduced, however, applications do not always follow.
Therefore, for the problem at hand, it is recommended to run the procedure several times   
before making the final decision of the selected fittest.  

The specific application of GA to graphical model search consists in how to encode
graphs as binary strings and in defining the operators selection, crossover and mutation.

Firstly, an undirected graph $G=(V,E)$ with $k$ vertices can be totally represented by 
a (strictly) upper triangular matrix $\M(G)=(m_{ij})_{j>i}$  
in which $m_{ij}=1$ iff there is an edge between the $i$-th and $j$-th vertices.
The matrix $\M(G)$ in turn can be identified with a binary string $\B(G)$
in which the entry $m_{ij}$ of $\M$ is stored at the corresponding position
$(k-1)(i-1)+(j-1)-i(i-1)/2$ of $\B$.  
For example, the true model with formula 12/23 in the previous example
can be encoded by the binary string $(1,0,1)$.
The length of binary strings encoding graphs with $k$ vertices is $k(k-1)/2$. 

The fitness function is inversely proportional to the loss rank:
the fitter an individual, the smaller its loss rank.
The fitness proportionate selection is not suitable in the present context,
because some loss ranks may be very close or even equal to zero,
which may cause premature convergence.   
Therefore the {\it linear ranking selection} \citep{Mitchell:96} should be used.
This selection operator starts by sorting the individuals in the decreasing (equivalently, increasing) order of
fitness (loss rank).
Then the probability for the $i$-th individual in the ranking to be selected is
\beqn
p_i=\fr1n\left(\b-2(\b-1)\fr{i-1}{n-1}\right),\ \ \b\in[1,2].
\eeqn
It seems that there is no clear suggestion on selection of $\b$.
In our simulation, $\b$ is fixed to $1.5$. 
We also apply the elitism in which $5\%$ of the fittest individuals are kept for the next population
before applying the selection operator. 

We now follow \cite{Poli:98} to define the crossover operator.
For a pair of parents models $G^1$ and $G^2$,
a subset $A$ of $V$ is randomly selected and two offspring are formed by exchanging the induced subgraphs $G_A^1, G_A^2$.
The motivation of this operator is interpreted in \cite{Poli:98}.
The mutation operator consists in randomly selecting a bit in a binary string and 
change its value (0 to 1 and vice versa).
The probability of doing crossover and mutation are fixed to $p_c=.9$ and $p_m=0.01$ respectively.
These values are chosen based on our own experience and in reference to others  

Some authors regard model selection as more than a machine learning or statistical issue,
it is a philosophical one!
Whether or not the true model exists is a controversial issue;
another one is that whether or not one should select a single model 
and do subsequent inferences conditional on that selected one.    
It would be risky to select a single model,
especially out of thousands as in the graphical modelling,
and proceed as if it was the true one. 
Even if the true model exists, it is unrealistic to expect the GA
to be always able to find.
It is therefore more reasonable to restrict our expectation to finding a set of appropriate models
instead of a single ``best" one (which may turn out to be an inappropriate model!).
The selected set then serves as the basic for a further context-specific consideration.  
The idea of selecting a set of models instead of a single model has also been discussed in \cite{Roverato:04}.

We now present the algorithm formally for searching for a set $\H$
of appropriate models.
The maximum cardinality of such a set is pre-specified, say $K$.
The basic idea is to repeat the GA procedure several times with different initial populations.
We start with $\H=\emptyset$.
After each iteration (of the GA procedure), a fittest individual is selected.
This individual will be added to the optimal set $\H$ if it was not previously selected.
The overall procedure stops when either the cardinality of $\H$ reaches $K$
or $\H$ does not change for the last, say $J$, iterations.   
The following is the GA-LR pseudo-code for our procedure.
(the readers who are not familiar with GA are referred to \cite{Mitchell:96} for the terminology).

\begin{list}{}{\parskip=0ex\parsep=0ex\itemsep=0.5ex\leftmargin=0ex\labelwidth=0ex}
  \item {\bf\boldmath The GA-LR algorithm}
  \begin{list}{}{\parskip=0ex\parsep=0ex\itemsep=0.5ex\leftmargin=2ex\labelwidth=1ex\labelsep=1ex}
    \item $\H:=\emptyset$, $j:=0$, resampling $B$ resamples
    \item[$\lceil$] {\bf While} $|\H|\leq K$ and $j\leq J$ {\bf do}	
    \begin{list}{}{\parskip=0ex\parsep=0ex\itemsep=0.5ex\leftmargin=2ex\labelwidth=1ex\labelsep=1ex}
      \item Generate an initial population $P$
      \item Calculate the loss ranks for models in $P$ and select the fittest $G^*$
      \item $t:=0$ 
	\begin{list}{}{\parskip=0ex\parsep=0ex\itemsep=0.5ex\leftmargin=2ex\labelwidth=1ex\labelsep=1ex}
        \item[$\lceil$] {\bf While} $t\leq T$ {\bf do}
		\begin{list}{}{\parskip=0ex\parsep=0ex\itemsep=0.5ex\leftmargin=2ex\labelwidth=1ex\labelsep=1ex}
		\item apply elitism and selection on $P$ to form new population $P_1$
		\item apply crossover on $P_1$ to form $P_2$
		\item apply mutation on $P_2$ to form $P_3$
		\item $P:=P_3$
		\item Calculate the loss ranks for $P$ and select the fittest $G'$
		\item {\bf If} $G'=G^*$ {\bf then} $t:=t+1$ {\bf else} $G^*:=G';\ t:=0$
		\end{list}
	\item [$\lfloor$] {\bf end while}
      \end{list}
    \item {\bf If} $G^*\in \H$ {\bf then} $j:=j+1$ {\bf else} $\H=\H\cup\{G^*\};\ j:=0$	
    \end{list}
    \item [$\lfloor$] {\bf end while}
  \end{list}
\end{list} 
\paradot{A simulation study}
We consider a moderate example with 6 vertices, each vertex takes on two values.
The total number of graphs is $2^{\binom{6}{2}}=32768$.
Datasets of size $n=10000$ are generated from the ``true" model $123/456$.
In our simulation, the parameters $T$ and $J$ are fixed to 5,
the size of initial populations is fixed to 100.
For a pre-specified maximum cardinality $K$, 
each run of the GA-LR algorithm produces a set $\H$ of optimal models in which $|\H|\leq K$.
We are interested in whether or not the selected set $\H$ contains the true model.
A small $K$ may be preferred because it eases the subsequent analysis,
but important models are more likely to be missed. 
We evaluate the performance of selection criteria (LR and BIC) in terms of proportions 
in which the selected set $\H$ covers the true model.
Table \ref{table3} shows those proportions over 10 replications for various $K$.
From the simulation results, we draw the following conclusions:
(i) The GA-BIC algorithm often terminates before the maximum $K$ is reached,
i.e., the GA-BIC is more stable than the GA-LR;
(ii) In contrast, the GA-BIC misses the true model more often than the GA-LR.   
An obvious drawback of the GA-LR is its computational time.
In this simulation, each run of the GA-LR requires approximately 30 minutes
which is about 50 times more than the running time of the GA-BIC.
\begin{table}[ht]
\centering 
\begin{tabular}{c|ccc}
$K$&10&20&50\\
\hline
GA-LR&.3&.6&.8\\
GA-BIC&.4&.5&.5
\end{tabular}
\caption{Proportions of correct coverage for various $K$ over 10 replications} 
\label{table3}
\end{table}
  
\paradot{Remarks on computation aspects}
The implementation is written in R,
benefited from the R package {\it igraph} of Gabor Csardi.
The simulation was carried out on a CPU Intel 2.66GHz.
The software is freely available upon contacting the authors.

\section{Conclusion}\label{secConclusion}
We have presented in this paper our continuous investigation 
of the LoRP, a general-purpose principle for model selection.
The efficiency of the LoRP for model selection in classification
was shown theoretically and experimentally.
We also developed the LoRP for model selection in unsupervised learning settings
and studied it by a means of simulation. 

A fundamental question in model selection is that what kind of model one would like to select:
the true model (i.e., the model generating the data) or a useful model in some sense (often, in terms of prediction)
or a parsimonious model that fits the data not too bad.
The LoRP attempts to deal with the latter 
which is, by common consent, the most appealing one in the machine learning community.
Our objective in this paper is to draw the reader's attention to a new methodology
for model selection that seems to have a lot of potential, leading to a rich field.

\begin{thebibliography}{33}
\expandafter\ifx\csname natexlab\endcsname\relax\def\natexlab#1{#1}\fi
\expandafter\ifx\csname url\endcsname\relax
  \def\url#1{{\tt #1}}\fi

\bibitem[Akaike(1973)]{Akaike:73}
H.~Akaike.
\newblock Information theory and an extension of the maximum likelihood
  principle.
\newblock In {\em Proc. 2nd International Symposium on Information Theory},
  pages 267--281, Budapest, Hungary, 1973. Akademiai Kaid\'o.

\bibitem[Allen(1974)]{Allen:74}
D.~Allen.
\newblock The relationship between variable selection and data augmentation and
  a method for prediction.
\newblock {\em Technometrics}, 16:\penalty0 125--127, 1974.

\bibitem[Arlot(2008)]{Arlot:08}
S.~Arlot.
\newblock Model selection by resampling penalization.
\newblock {\em Electronic Journal Statist.}, 2008.

\bibitem[Bartlett et~al.(2002)Bartlett, Boucheron, and Lugosi]{Bartlett:02}
P.~Bartlett, S.~Boucheron, and G.~Lugosi.
\newblock Model selection and error estimation.
\newblock {\em Machine Learning}, 48:\penalty0 85--113, 2002.

\bibitem[Bartlett et~al.(2005)Bartlett, Bousquet, and Mendelson]{Bartlett:05}
P.~L. Bartlett, O.~Bousquet, and S.~Mendelson.
\newblock Local rademacher complexities.
\newblock {\em Ann. Statist.}, 33\penalty0 (4):\penalty0 1497--1537, 2005.

\bibitem[Calinski and Harabasz(1974)]{Calinski:74}
R.~B. Calinski and J.~Harabasz.
\newblock A dendrite method for cluster analysis.
\newblock {\em Communications in Statistics}, 3:\penalty0 1--27, 1974.

\bibitem[Craven and Wahba(1979)]{Craven:79}
P.~Craven and G.~Wahba.
\newblock Smoothing noisy data with spline functions: estimating the correct
  degree of smoothing by the methods of generalized cross-validation.
\newblock {\em Numerische Mathematik}, 31:\penalty0 377--403, 1979.

\bibitem[Edwards(2000)]{Edwards:00}
David Edwards.
\newblock {\em Introduction to Graphical Modelling}.
\newblock Springer-Verlay New York, 2000.
\newblock 2nd.

\bibitem[Efron and Tibshirani(1993)]{Efron:93}
B.~Efron and R.J. Tibshirani.
\newblock {\em An Introduction to the Bootstrap}.
\newblock Chapman \& Hall, 1993.

\bibitem[Fan and Li(2001)]{Fan:01}
J.~Fan and R.~Li.
\newblock Variable selection via nonconcave penalized likelihood and its oracle
  properties.
\newblock {\em JASA}, 96\penalty0 (456):\penalty0 1348--1360, 2001.

\bibitem[Fromont(2007)]{Fromont:07}
M.~Fromont.
\newblock Model selection by bootstrap penalization for classification.
\newblock {\em Mach. Learn.}, 66:\penalty0 165--207, 2007.

\bibitem[Gine and Zinn(1990)]{Gine:90}
E.~Gine and J.~Zinn.
\newblock Bootstrapping general empirical functions.
\newblock {\em Ann. Probab..}, 18:\penalty0 851--869, 1990.

\bibitem[Hastie et~al.(2005)Hastie, Tibshirani, and Friedman]{Hastie:05}
T.~Hastie, R.~Tibshirani, and J.~H. Friedman.
\newblock {\em The Elements of Statistical Learning}.
\newblock Springer, 2005.

\bibitem[Holland(1975)]{Holland:75}
J.~H. Holland.
\newblock {\em Adaption in Natural and Artificial Systems}.
\newblock University of Michigan Press, 1975.

\bibitem[Hutter(2007)]{Hutter:07}
M.~Hutter.
\newblock The loss rank principle for model selection.
\newblock In {\em Proc. 20th Annual Conf. on Learning Theory ({COLT'07})},
  volume 4539 of {\em LNAI}, pages 589--603, San Diego, 2007. Springer, Berlin.
\newblock URL \url{http://arxiv.org/abs/math.ST/0702804}.

\bibitem[Hutter and Tran(2010)]{Hutter:10}
M.~Hutter and M.-N. Tran.
\newblock Model selection with the loss rank principle.
\newblock {\em Computational Statistics and Data Analysis}, 54\penalty0
  (5):\penalty0 1288--1306, 2010.

\bibitem[Koltchinskii(2001)]{Koltchinskii:01}
V.~Koltchinskii.
\newblock Rademacher penalties and structural risk minimization.
\newblock {\em IEEE Trans. Inform. Theory}, 47:\penalty0 1902--1914, 2001.

\bibitem[Koltchinskii(2006)]{Koltchinskii:06}
V.~Koltchinskii.
\newblock Local rademacher complexities and oracle inequalities in risk
  minimization.
\newblock {\em Ann. Statist.}, 34\penalty0 (6):\penalty0 2593--2656, 2006.

\bibitem[Lozano(2000)]{Lozano:00}
F.~Lozano.
\newblock Model selection using rademacher penalization.
\newblock In {\em Proc. 2nd ICSC Symp. Neural Computation NC2000}. Berlin,
  Germany: ICSC Academic, 2000.

\bibitem[Mallows(1973)]{Mallows:73}
C.~L. Mallows.
\newblock Some comments on $c_p$.
\newblock {\em Technometrics}, 15\penalty0 (4):\penalty0 661--675, 1973.

\bibitem[Mitchell(1996)]{Mitchell:96}
M.~Mitchell.
\newblock {\em An Introduction to Genetic Algorithms}.
\newblock The MIT Press, 1996.

\bibitem[Poli and Roverato(1998)]{Poli:98}
I.~Poli and A.~Roverato.
\newblock A genetic algorithm for graphical model selection.
\newblock {\em Journal of the Italian Statistical Society}, 7\penalty0
  (2):\penalty0 197--208, 1998.

\bibitem[Rissanen(1978)]{Rissanen:78}
J.~J. Rissanen.
\newblock Modeling by shortest data description.
\newblock {\em Automatica}, 14\penalty0 (5):\penalty0 465--471, 1978.

\bibitem[Roverato and Paterlini(2004)]{Roverato:04}
A.~Roverato and S.~Paterlini.
\newblock Technological modelling for graphical models: an approach basedon
  genetic algorithms.
\newblock {\em Computational Statistics \& Data Analysis}, 47:\penalty0
  323--337, 2004.

\bibitem[Schwarz(1978)]{Schwarz:78}
G.~Schwarz.
\newblock Estimating the dimension of a model.
\newblock {\em Annals of Statistics}, 6\penalty0 (2):\penalty0 461--464, 1978.

\bibitem[Shao(1996)]{Shao:96}
J.~Shao.
\newblock Bootstrap model selection.
\newblock {\em Journal of the American Statistical Association}, 91\penalty0
  (434):\penalty0 655--665, 1996.

\bibitem[Shibita(1984)]{Shibita:84}
R.~Shibita.
\newblock Approximate efficiency of a selection procedure for the number of
  regression variables.
\newblock {\em Biometrika}, 71:\penalty0 43--49, 1984.

\bibitem[Tibshirani(1996)]{Tibshirani:96}
R.~Tibshirani.
\newblock Regression shrinkage and selection via the lasso.
\newblock {\em J. R. Statist. Soc. B}, 58\penalty0 (1):\penalty0 267--288,
  1996.

\bibitem[Tran(2009)]{Tran:09}
M.~N. Tran.
\newblock Penalized maximum likelihood principle for choosing ridge parameter.
\newblock {\em Communications in Statistics - Simulation and Computation},
  38:\penalty0 1610--1624, 2009.

\bibitem[Tran(2010)]{Tran:10}
M.~N. Tran.
\newblock The loss rank criterion for variable selection in linear regression
  analysis.
\newblock {\em Scandinavian Journal of Statistics}, 2010.
\newblock to appear.

\bibitem[van~der Vaart and Wellner(1996)]{vdV:96}
A.~W. van~der Vaart and J.~A. Wellner.
\newblock {\em Weak convergence and empirical processes}.
\newblock Springer, 1996.

\bibitem[Vapnik and Chervonenkis(1971)]{Vapnik:71}
V.~N. Vapnik and A.Y. Chervonenkis.
\newblock On the uniform convergence of relative frequencies of events to their
  probabilities.
\newblock {\em Theory Prob. its Application}, 16:\penalty0 264--280, 1971.

\bibitem[Whittaker(1990)]{Whittaker:90}
J.~Whittaker.
\newblock {\em Graphical Models in Applied Multivariate Statistics}.
\newblock Wiley, 1990.

\end{thebibliography}

\end{document}